\documentclass[prl,twocolumn,english,superscriptaddress,footinbib]{revtex4-1}

\usepackage{amsmath,amssymb,mathrsfs}
\usepackage{amsfonts,bm}
\usepackage{amsmath}
\usepackage{amssymb}
\usepackage{amsthm}
\usepackage{graphicx}
\usepackage{graphics}
\usepackage{subfigure}
\usepackage{color}
\usepackage{bm}
\usepackage{hyperref}
\usepackage{rotating}
\usepackage{comment}
\usepackage[colorinlistoftodos]{todonotes}

\usepackage{xfrac}

\newcommand{\be}{\begin{equation} }
\newcommand{\ee}{\end{equation} }
\newcommand{\ba}{\begin{eqnarray} }
\newcommand{\ea}{\end{eqnarray} }

\newcommand{\bpm}{\begin{pmatrix}}
\newcommand{\epm}{\end{pmatrix}}
\newcommand{\bmm}{\begin{matrix}}
\newcommand{\emm}{\end{matrix}}

\newcommand{\la}{\label}
\newcommand{\p}{\partial}
\newcommand{\sech}{\text{sech}}
\newcommand{\bea}{\begin{eqnarray}}
\newcommand{\eea}{\end{eqnarray}}

\makeatother

\usepackage{babel}

\begin{document}

\title{Non-linear shallow water dynamics with odd viscosity}

\author{Gustavo M. Monteiro}
 \author{Sriram Ganeshan} 
 \affiliation{Department of Physics, City College, City University of New York, New York, NY 10031, USA }

\date{\today}

\begin{abstract}
In this letter, we derive the Korteweg-de Vries (KdV) equation corresponding to the surface dynamics of a shallow depth ($h$) two-dimensional fluid with odd viscosity ($\nu_o$) subject to gravity ($g$) in the long wavelength  weakly nonlinear limit. In the long wavelength limit, the odd viscosity term plays the role of surface tension albeit with opposite signs for the right and left movers. We show that there exists two regimes with a sharp transition point within the applicability of the KdV dynamics, which we refer to as weak $(|\nu_o|< \sqrt{gh^3}/6)$ and strong $(|\nu_o|> \sqrt{gh^3}/6)$ parity-breaking regimes. While the `weak' parity breaking regime results in minor qualitative differences in the soliton amplitude and velocity between the right and left movers, the `strong' parity breaking regime on the contrary results in  solitons of depression (negative amplitude) in one of the chiral sectors. 
\end{abstract}
\maketitle

\textit{\textbf{Introduction:}} Parity-breaking phenomena in two-dimensional fluids such as odd viscosity effects have been at the center of investigation in diverse platforms. For two dimensional isotropic and non-dissipative flows, odd viscosity is the only non-vanishing coefficient of the viscosity tensor and has a geometrical connection to the adiabatic curvature on the space of flat background metrics~\cite{avron1995viscosity, avron1998odd}. Examples of quantum systems where odd viscous effects are important include electron fluids in mesoscopic systems~\cite{scaffidi2017hydrodynamic, pellegrino2017nonlocal, berdyugin2019measuring}, quantum Hall fluids~\cite{tokatly2006magnetoelasticity,tokatly2007new,tokatly2009erratum, read2009non,haldane2011geometrical,haldane2011self,hoyos2012hall, bradlyn2012kubo, yang2012band,abanov2013effective,hughes2013torsional, hoyos2014hall, laskin2015collective, can2014fractional,can2015geometry,klevtsov2015geometric,klevtsov2015quantum, gromov2014density, gromov2015framing, gromov2016boundary,alekseev2016negative},  and chiral superfluids/superconductors~\cite{hoyos2014effective}.  

In classical fluids, odd viscosity shows up in polyatomic gases~\cite{korving1966transverse, knaap1967heat, korving1967influence, hulsman1970transverse}, chiral active matter \cite{banerjee2017odd, souslov2019topological, soni2018free}, vortex dynamics in 2D~\cite{wiegmann2014anomalous, yu2017emergent, bogatskiy2018edge, bogatskiy2019vortex} and chiral active fluids~\cite{banerjee2017odd, souslov2019topological, soni2018free}. For incompressible flows, it has been shown by one of the authors that odd viscosity effects are absent when the fluid is spread on the entire plane or confined in rigid domains with no-slip boundary conditions \cite{ganeshan2017odd}. In other words, the velocity profile is independent of the odd viscosity. Nevertheless, the signature of this parity-breaking coefficient is present in surface waves and in the interface between two fluids governed by kinematic and no-stress boundary conditions, which explicitly depends on the odd viscosity. The dynamical surface problem in the presence of odd viscosity results in an oscillating boundary layer where the vorticity is confined within some thickness of $\delta \propto \sqrt\nu_e$~\cite{abanov2018odd, abanov2018free} (where $\nu_e$ is the kinematic shear viscosity) for the dissipative case and $\delta \propto c_s^{-1}$ (where $c_s$ is the sound velocity) for the non-dissipative compressible case~\cite{abanov2019hydro}. In the limit of a very thin boundary layer, that is, $\nu_e\rightarrow 0$ or $c_s\rightarrow \infty$, both the fluid pressure at the edge and the surface vorticity diverge as $ 1/\sqrt{\nu_e}$ or $ c_s$, but the quantity $\tilde p=p-\nu_o\rho\,\omega$ remains finite. We refer to $\tilde p$ as modified pressure, where $\nu_o$ is the odd viscosity and the variables $p$, $\omega$ and $\rho$ are the fluid pressure, vorticity and constant background density respectively. This cancelation of divergences allows us to write the dynamical surface problem with odd viscosity as an effective irrotational system where all the effects of odd viscosity and boundary layer can be absorbed into a modified pressure term at the edge.

In short, for an irrotational flow, that is, $\boldsymbol v=\boldsymbol\nabla\theta$, the effective boundary dynamics can be expressed as a Laplace equation for the velocity potential $\theta$ in the bulk and Bernoulli's equation at the boundary with the modified pressure~\cite{abanov2018odd}. The variational principle for this boundary dynamics was later formulated in terms of a geometric action which resulted in the  odd viscosity induced effective pressure at the boundary~\cite{abanov2018free}. In the limit of infinitely deep fluid the weakly non-linear dynamics within a small angle approximation was shown to be governed by the novel {\it Chiral-Burgers} equation~\cite{abanov2018odd}. 

\begin{figure}
\centering
\includegraphics[scale=0.28]{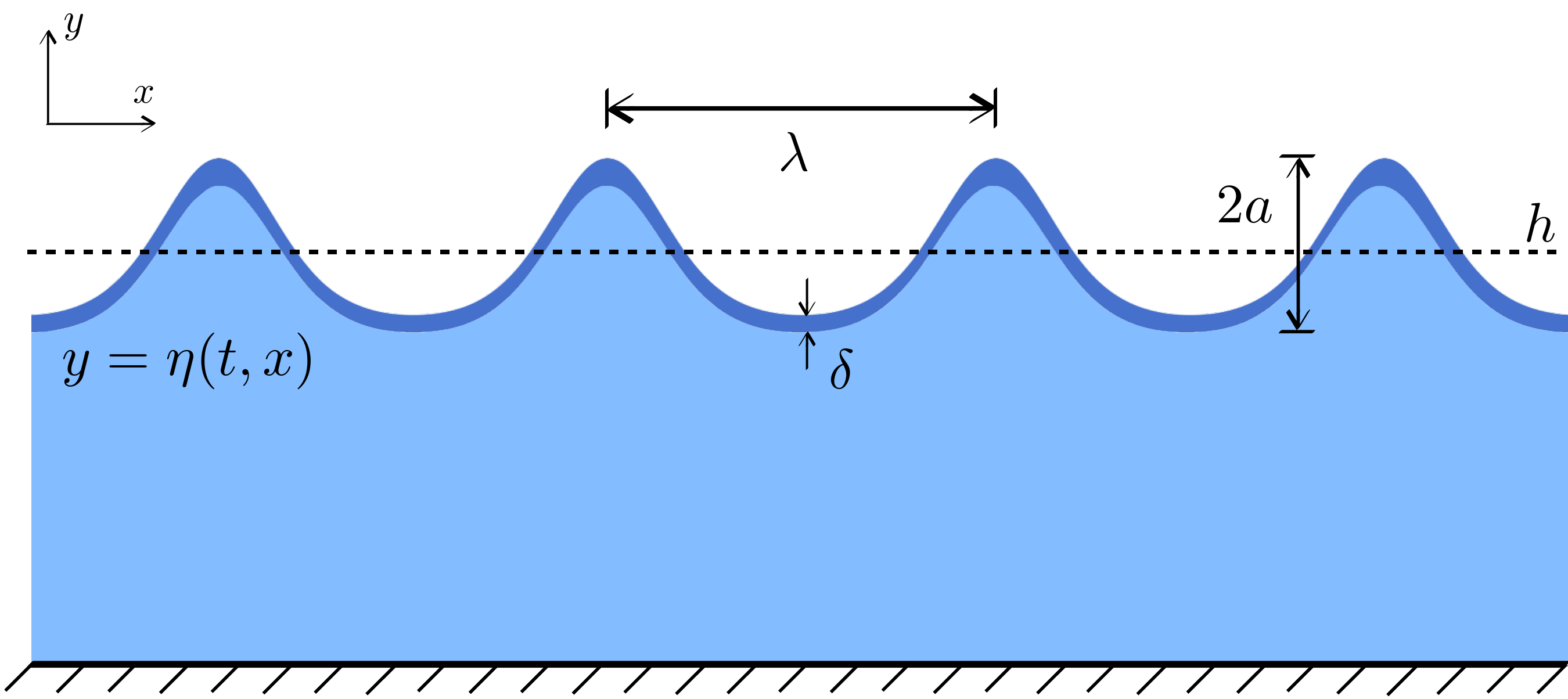}
\caption{Schematics of the shallow fluid dynamics with free surface. Here $a$ denotes the amplitude of the edge profile and $\delta$ is the boundary layer thickness. The vertical direction is exaggerated to highlight these features.}
\label{fig:schematic}
\end{figure}

In this letter, we study the shallow depth limit of the weakly non-linear surface dynamics with odd viscosity and gravitational force (confining potential) (see the schematic in Fig.~\ref{fig:schematic}). We assume that the boundary layer is the shortest length scale and the effective dynamics is irrotational, using the hydrodynamic equations from~\cite{abanov2018free} as the starting point. We show that, for later times and long wavelengths, the weakly non-linear dynamics is given by the integrable Kortweg De Vries (KdV) equation with the  kinematic odd viscosity $\nu_o$ entering the coefficient of the dispersive term, 
\begin{align}
	\eta_x\pm\frac{\eta_t}{\sqrt{gh}}+\frac{3}{2h}\eta\, \eta_x+h^2\left(\frac{1}{6}\pm \frac{\nu_o}{\sqrt{gh^3}}\right) \eta_{xxx}=0. \la{eq:eta-dim}
\end{align}
Here, $\eta(x,t)$ is the boundary shape profile, $h$ is the average depth of the fluid and $g$ is the acceleration of gravity. The positive sign refers to right-moving solitons whereas the negative sign refers to the left-moving ones. The manifestation of odd viscosity in the above equation is similar to that of the surface tension although with different signs for the left and right movers.

The odd viscosity entering the KdV equation has major consequences due to its parity breaking effects. We show that there exists two regimes with a sharp transition point within the applicability of the KdV dynamics, which we refer to as weak $(|\nu_o|< \sqrt{gh^3}/6)$ and strong $(|\nu_o|> \sqrt{gh^3}/6)$ parity-breaking regimes. In the weak parity breaking regime, the left and right moving solitons slightly differ in amplitude and speed. In the strong parity breaking regime, one of the sectors becomes solitonic waves of depression. At the critical value $(|\nu_o|=\sqrt{gh^3}/6)$, one of the sectors  becomes unstable and higher order derivative terms become important. The parity breaking KdV dynamics discussed here is in stark contrast to the parity preserving case of shallow water KdV dynamics without odd viscosity.

\textit{\textbf{Incompressible fluids with odd viscosity:}} The hydrodynamic equations for incompressible fluids with odd viscosity consist of the Euler equation, with modified pressure, together with the incompressibility condition, that is,
\be
\boldsymbol{\nabla\cdot v}=0\,,\quad \p_t\boldsymbol v+(\boldsymbol{v\cdot\nabla})\boldsymbol v=-\frac{1}{\rho}\boldsymbol\nabla\tilde p-g\boldsymbol{\hat y}\,.  \la{eq:euler}
\ee
Here, $\boldsymbol v$ is the flow velocity and $\rho$ is the constant and uniform fluid density. This differs from the ordinary Euler equation in the definition of the modified pressure
\be
\tilde p:=p-\nu_o\rho\,\omega\,,
\ee
where $\omega=\boldsymbol{\nabla}\times \boldsymbol{v}$ is the fluid vorticity.

Since the fluid density is constant, there is no equation of state and the pressure is completely determined by the flow. The curl of the Euler equation is an equation for the flow vorticity, which does not depend on the pressure. This vorticity equation together with the incompressibility condition completely determines the fluid flow, up to boundary conditions. Therefore, the presence of odd viscosity will only change the velocity flow if the boundary conditions depend on $\nu_o$, otherwise, it only modifies how the pressure depends on velocity flow, which is not an easily accessible quantity in experiments.

{\bf Irrotational limit of the free surface dynamics and boundary layer approximation:}
Bulk equations of motion~(\ref{eq:euler}) must be supplemented with boundary conditions. The fluid free surface is a dynamical interface $y=\eta(t,x)$ between two fluids where we impose one kinematic and two dynamical boundary conditions. The kinematic boundary condition states that the velocity of the fluid normal to the boundary is equal to the rate of change of the boundary shape. The pair of dynamical boundary conditions imposes that there are no normal and tangent forces acting on an element of the fluid surface. 

 The presence of two dynamical boundary conditions along with incompressibility requires a singular boundary layer where the vorticity is confined. The role of this singular boundary layer is to ensure that there are no tangent forces on this interface. It has been shown that regardless of the boundary layer mechanism (dissipative or compressible), the normal component of the DBC is universal and geometric in nature~\cite{abanov2018odd, abanov2018free, abanov2019hydro}. Assuming that the boundary layer is stable and is confined to short length scales, the free surface problem can be written as an effective description of the fluid with the effects of boundary layer encoded in the odd viscosity modified pressure term,
\begin{align}
	\tilde p\big|_{y=\eta(t,x)}= \frac{2\nu_o}{\sqrt{1+\eta_x^2}} \p_xv_n\,. \la{DBC}
\end{align} 
Here $v_n$ is the velocity component which is normal to the boundary, taken at $y=\eta(t,x)$. 

This effective free surface dynamics, that is, Eqs.~(\ref{eq:euler}, \ref{DBC}) can be expressed in the form of an action principle, as shown in \cite{abanov2018free}. For irrotational flows, i.e., $\boldsymbol v=\boldsymbol\nabla\theta$, this action simplifies and can be thought of as an odd viscosity extension of the Luke's variational principle \cite{luke1967variational}. Following \cite{abanov2018free}, the hydrodynamic action becomes
\begin{align}
	S =&\,-\iint dt\,dx \int_{-h}^{\eta(t,x)}dy\, \left[\theta_t+\frac{1}{2}\left(\theta_x^2+\theta_y^2\right) \right] \nonumber
	\\
	&-\iint dt\, dx\left[\frac{1}{2}g\eta^2-\nu_o\,\eta_t\tan^{-1}(\eta_x)\right],
 \label{eq:LVP}
\end{align}
where $\eta(t,x)$ as the top surface shape function. The domain of the fluid is bounded by a finite depth at the bottom. The bulk and boundary equations of motion are obtained by varying the above action for a finite depth fluid. The bulk equation for the irrotational system is simply the Laplace equation for the potential $\Delta \theta=0$ defined in the domain $-h<y<\eta(t,x)$. The boundary conditions at the top and bottom of the fluid domain can be written as,
\begin{align}
       &\theta_y=0\,, \qquad \qquad \qquad \qquad  y=-h, \la{eq:bottomBC}
    \\
    &\eta_t+\eta_x\theta_x=\theta_y\,, \qquad \qquad y=\eta(t,x), \la{eq:KBC}
    \\
    &\theta_t+\frac{\theta_x^2+\theta_y^2}{2}+g\eta=\frac{2\nu_o}{\sqrt{1+\eta_x^2}}\,\p_x\left[\frac{\eta_t}{\sqrt{1+\eta_x^2}}\right],  \;\, y=\eta(t,x). \la{eq:DBC}
\end{align}
Eq.~(\ref{eq:KBC}) is the kinematic boundary condition and Eq.~(\ref{eq:DBC}) is the odd viscosity modified dynamic boundary condition.

{\bf Linear waves in long wavelength limit:} Before we dive into the non-linear dynamics of the shallow fluid regime, let us focus on the linearized free surface problem with finite depth. In this limit, we drop all the quadratic terms in Eqs.~(\ref{eq:KBC}, \ref{eq:DBC}) and evaluate the derivatives of $\theta$ at $y=0$. For the monochromatic surface profile $\eta(x,t)=a \cos(kx-\Omega t)$ as an input, we find the velocity potential to be 
\[
\theta(x,y,t)= \frac{a\,\Omega }{k\sinh(kh)}  \cosh[k(y+h)]\sin(kx-\Omega t),
\] 
with the surface dispersion relation given by
\begin{equation}
\Omega=\tanh kh\left[-\nu_o k^2\pm \sqrt{\nu_o^2 k^4+g k\coth (kh)}\right]. \la{Omega}
\end{equation}
In the absence of gravity $(g=0)$ and in the deep ocean limit $(h\rightarrow \infty)$, we have that $\tanh kh \approx k/|k|$ and we recover the known odd viscosity dominated dispersion $\Omega=\{0,-2\nu_ok|k|\}$. The weakly non-linear dynamics for this system was discussed in Ref.~\cite{abanov2018odd}. 

Shallow waves, on the other hand, arise when the fluid depth $h$ is much smaller than the characteristic wavelengths of the system. In other words, they are characterized by $kh\ll1$. In this approximation, the leading terms in the dispersion (\ref{Omega}) are given by
\begin{align}
	\Omega\approx \sqrt{\frac{g}{h}}\left[\pm  kh -\left(\frac{\nu_o}{\sqrt{gh^3}}\pm\frac{1}{6}\right)(kh)^3\right]   . \la{Omega-KdV}
\end{align}
The first term is the usual shallow water gravity wave dispersion, whereas the second one is the odd viscosity modified Korteweg de-Vries dispersive term. {\it Prima facie} it seems that the odd viscosity is qualitatively similar to the surface tension effect for the shallow water surface dispersion.  However, the physical manifestation of odd viscosity is completely different, since the coefficient of the cubic term can develop a relative sign change between the left mover and right mover for $|\nu_o|>\tfrac{1}{6}\sqrt{gh^3}$, what indicates a strong parity-breaking phenomena.

The shallow wave condition naturally introduces a power expansion in $kh$. Formally, it is convenient to define an expansion parameter $\varepsilon\ll 1$, such that $kh=\sqrt{\varepsilon}\, \bar k$ and $\bar k$ is a dimensionless wavenumber. Since an expansion in powers of $kh$ can be translated into a derivative expansion for the fluid dynamics, this rescaling is equivalent to the redefinition $x=\frac{h}{\sqrt\varepsilon}X$, where $X$ is the dimensionless horizontal coordinate. In the same way, we can define $y=hY$, with $Y$ being the dimensionless vertical coordinate. In this counting scheme, we have that $\p_x\sim \mathcal O\left(\varepsilon^{1/2}\right)$, whereas $\p_y\sim \mathcal O(1)$.

The wave dynamics dictated by Eq.~(\ref{Omega-KdV}) evolves according to two distinct time scales. The linear term scales with $\sqrt{\varepsilon}$ and governs the splitting of an initial disturbance into right-moving and left-moving wavepackets. The first time scale, which we denote by $T$, is the characteristic time in which left and right-movers are so far apart, we can study them separately. In other words, for sufficiently later times, the only role of $T$ is to account for the boost of the center of mass and it only appears in the combination $X-T$, for right-movers, or $X+T$, for left-movers.  On the other hand, the cubic term scales as $\varepsilon^{3/2}$ and give rise to a dispersive group velocity of the boosted wavepacket. This effect becomes relevant at much later times, in comparison to $T$, and introduce a second time scale, which governs the time evolution of the boosted wavepacket and we denote by $\tau$. This means that both variables $\theta$ and $\eta$ evolve according to this double time scale, such that, the time derivative becomes
\begin{equation*}
\p_t=\sqrt{\frac{g\varepsilon}{h}}\p_{T}+\sqrt{\frac{g\varepsilon^3}{h}}\p_\tau\,. \la{time-scales}
\end{equation*}

In the following, we derive the full non-linear shallow water dynamics with odd viscosity and discuss how the parity breaking effects manifest in the non-linear dynamics. In particular we show that  $\nu_o=\tfrac{1}{6}\sqrt{gh^3}$ manifests as a critical point that separates two qualitatively different regimes of non-linear dynamics.

{\bf Non-linear shallow depth waves:}
Korteweg-€"de Vries (KdV) equation arises in the study of shallow water waves of long wavelengths and small amplitudes. In the following analysis, we show how the KdV equation corresponding to the shallow depth limit is modified by the presence of the odd viscosity term. The counting scheme for the KdV equation is chosen such that small amplitudes regime corresponds to $\eta\sim\mathcal O (\varepsilon)$, that is, $\eta$ is of the same order as $\p_x^2$. Thus, we can rescale the boundary shape in terms of $h$ as $\eta=\varepsilon h\mathfrak y$. 

KdV regime happens for sufficiently later times, so that right-moving and left-moving solutions are independent and well-separated. Here, we restrict ourselves to only right-moving propagation, since the analysis for the the left-movers follows similarly. Hence, let us assume $\theta$ and $\eta$ of the form
\begin{align}
    \theta(t,x,y)&= \sqrt{\varepsilon h^3g}\;\vartheta(\tau,\sigma, Y;\varepsilon)\,,
    \\
    \eta(t,x)&=\varepsilon h\;\mathfrak y(\tau,\sigma;\varepsilon)\,,
\end{align}
with $\sigma=X-T$. Under these conditions, the bulk equation of motion and the boundary condition at the flat bottom become 
\begin{align}
    \varepsilon\,\vartheta_{\sigma\sigma}+\vartheta_{Y Y}&=0, \qquad -1<Y<\varepsilon\mathfrak y, \la{eq:harmonic}
    \\
    \vartheta_{Y}&=0, \qquad\qquad Y=-1. \la{eq:bottomBC2}
\end{align}

Let us denote $\vartheta(\tau,\sigma, -1;\varepsilon)$ by $\phi(\tau,\sigma;\varepsilon)$. This way, the solution of Eq.~(\ref{eq:harmonic}) with the condition (\ref{eq:bottomBC2}) can be written as
\begin{equation}
    \vartheta(\tau,\sigma, Y; \varepsilon)=\sum_{n=0}^\infty \frac{(-\varepsilon)^n(1+Y)^{2n}}{(2n)!}\,\p_{\sigma}^{2n}\phi\,. \la{eq:potential}
\end{equation}

Plugging Eq.~(\ref{eq:potential}) into Eqs.~(\ref{eq:KBC}, \ref{eq:DBC}) and neglecting terms of $\mathcal O(\varepsilon^2)$ or higher, we obtain 
\begin{align}
   \phi_\sigma &=\mathfrak y+\varepsilon\left(\phi_\tau+\tfrac{1}{2}\phi_\sigma^2+\tfrac{1}{2}\phi_{\sigma\sigma\sigma}+2\bar\nu_o\mathfrak y_{\sigma\sigma}\right), \la{eq:phi}
    \\
    \mathfrak y_\sigma-\phi_{\sigma\sigma}&= \varepsilon\left[\tfrac{1}{2}\mathfrak y_{\sigma\sigma\sigma}-\mathfrak y_\tau+\tfrac{2}{3}\phi_{\sigma\sigma\sigma\sigma}+\p_\sigma(\phi_\sigma\mathfrak y)\right]. \la{eq:eta}
\end{align}
Here, we denoted $\bar\nu_o=\nu_o/\sqrt{gh^3}$. Eq.~(\ref{eq:phi}) allow us to perturbatively express $\phi_\sigma$ in terms of $\mathfrak y$. Substituting this expression into Eq.~(\ref{eq:eta}), the leading order equation for the right-moving surface wave in the boosted reference frame becomes
\begin{equation}
    \mathfrak y_\tau+\tfrac{3}{2}\mathfrak y\,\mathfrak y_\sigma+\left(\tfrac{1}{6}+\bar\nu_o\right)\mathfrak y_{\sigma\sigma\sigma}=0, \la{eq:right-mov}
\end{equation}
which is nothing but the well-known KdV equation. In terms of the dimensionful variables $t$, $x$ and $\eta(t,x)$, it becomes Eq.~(\ref{eq:eta-dim}).

As previously mentioned, we could repeat the same analysis for the left-moving solitons. For $\xi=X+T$, we obtain
\begin{equation}
     \mathfrak y_\tau-\tfrac{3}{2}\mathfrak y\,\mathfrak y_\xi-\left(\tfrac{1}{6}-\bar\nu_o\right)\mathfrak y_{\xi\xi\xi}=0. \la{eq:left-mov}
\end{equation}
Under reflection about y-axis (parity operation), $\mathfrak y\rightarrow\mathfrak y$,  $\xi\rightarrow-\sigma$ and Eq.~(\ref{eq:left-mov}) becomes
\begin{equation}
    \mathfrak y_\tau+\tfrac{3}{2}\mathfrak y\,\mathfrak y_\sigma+\left(\tfrac{1}{6}-\bar\nu_o\right)\mathfrak y_{\sigma\sigma\sigma}=0. \la{eq:left-mov-T}
\end{equation}

The odd viscosity term breaks parity symmetry of the problem \footnote{A right moving soliton reflected about Y-axis is same as this solution played backwards in time.}, since the left-moving soliton under reflection about the y-axis do not behave like the right-moving soliton. The odd viscosity term entering the KdV equation is similar to the presence of the surface tension. Within this analogy of odd viscosity as surface tension, the left mover and right mover will have opposite signs of surface tension due to the parity breaking effects of odd viscosity. 
In other words, odd viscosity in the KdV regime acts as chirality dependent surface tension term. 

{\bf Soliton solution:} In the following we analyze the role of odd viscosity in the single soliton solution of Eqs.~(\ref{eq:right-mov}) and (\ref{eq:left-mov}). Although multi-soliton solutions also show the same qualitative behavior, they are out of the scope of this letter. The single soliton solution corresponding to the left and right movers can be written as,
\begin{equation}
\mathfrak y(\tau,x_\pm)=8\bar k^2\left(\tfrac{1}{6}\mp\bar\nu_o\right)\text{sech}^2\left[\bar k x_\pm\pm4\bar k^3\tau\left(\tfrac{1}{6}\mp\bar\nu_o\right)\right],
\end{equation}
where $\bar k$ is the dimensionless wavenumber and we denoted $x_+=\xi$ and $x_-=\sigma$ in order to shorten the notation. Moreover, $\tau=0$ was chosen such that the soliton center of mass is at $x_\pm=0$. Note that $\left(\tfrac{1}{6}\mp\bar\nu_o\right)$ enters both the amplitude and the wave speed. Therefore, the odd viscosity modification to the KdV soliton dynamics can be separated into three regimes depending on the value of $\left(\tfrac{1}{6}\mp\bar\nu_o\right)$

\begin{figure}
\centering
\includegraphics[scale=0.7]{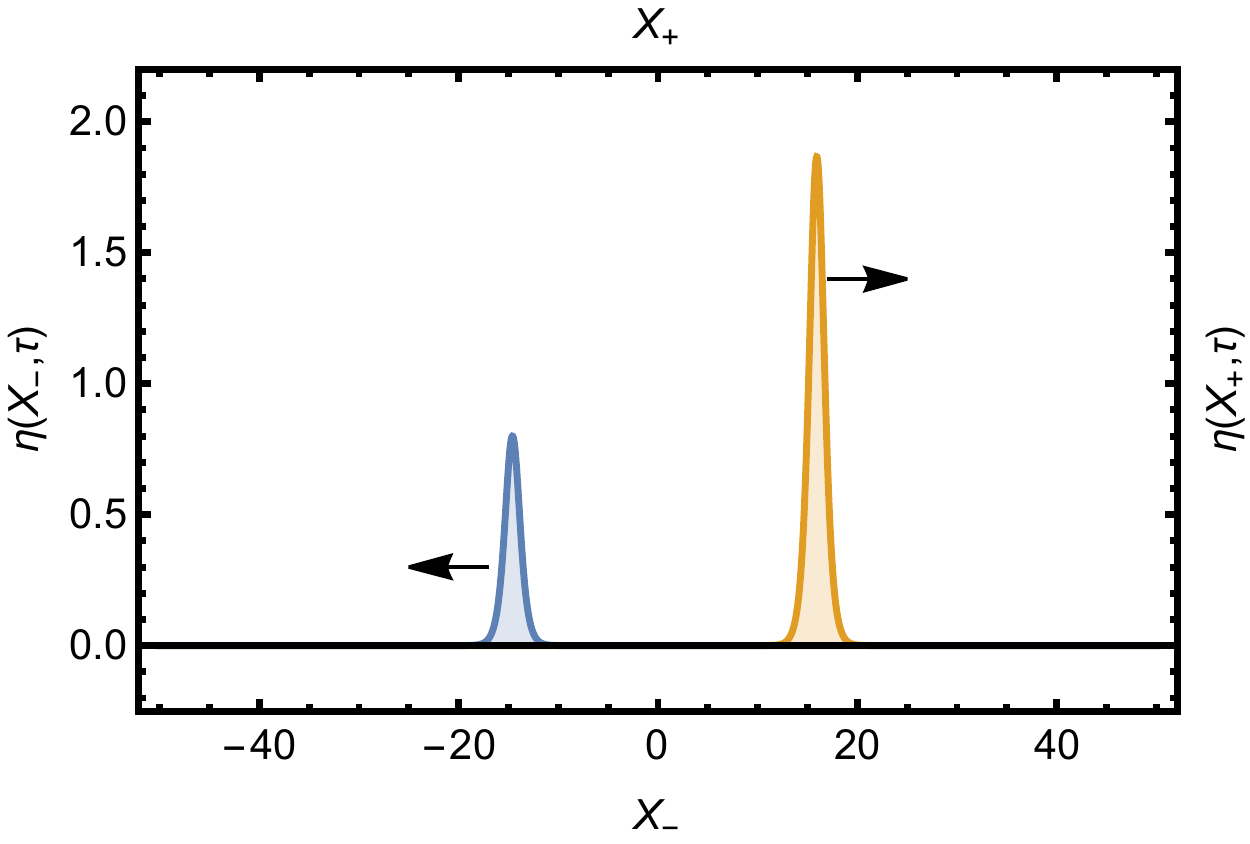}
\caption{Left (blue) and right (yellow) moving soliton in the weak parity breaking regime with parameters $|\bar \nu_o|=\tfrac{1}{15}$, $\bar k=1$ and $\tau=1$.}
\label{fig:wp}
\end{figure}

{\it `Weak' parity breaking regime  $|\bar \nu_o|<\tfrac{1}{6}$:} In this case, $\left(\tfrac{1}{6}\mp\bar\nu_o\right)>0$, with left and right moving solitons only differ in the magnitude of the amplitude and velocity as shown in Fig.~\ref{fig:wp} . We refer to this as {\it weak parity breaking regime}.

\begin{figure}
\centering
\includegraphics[scale=0.7]{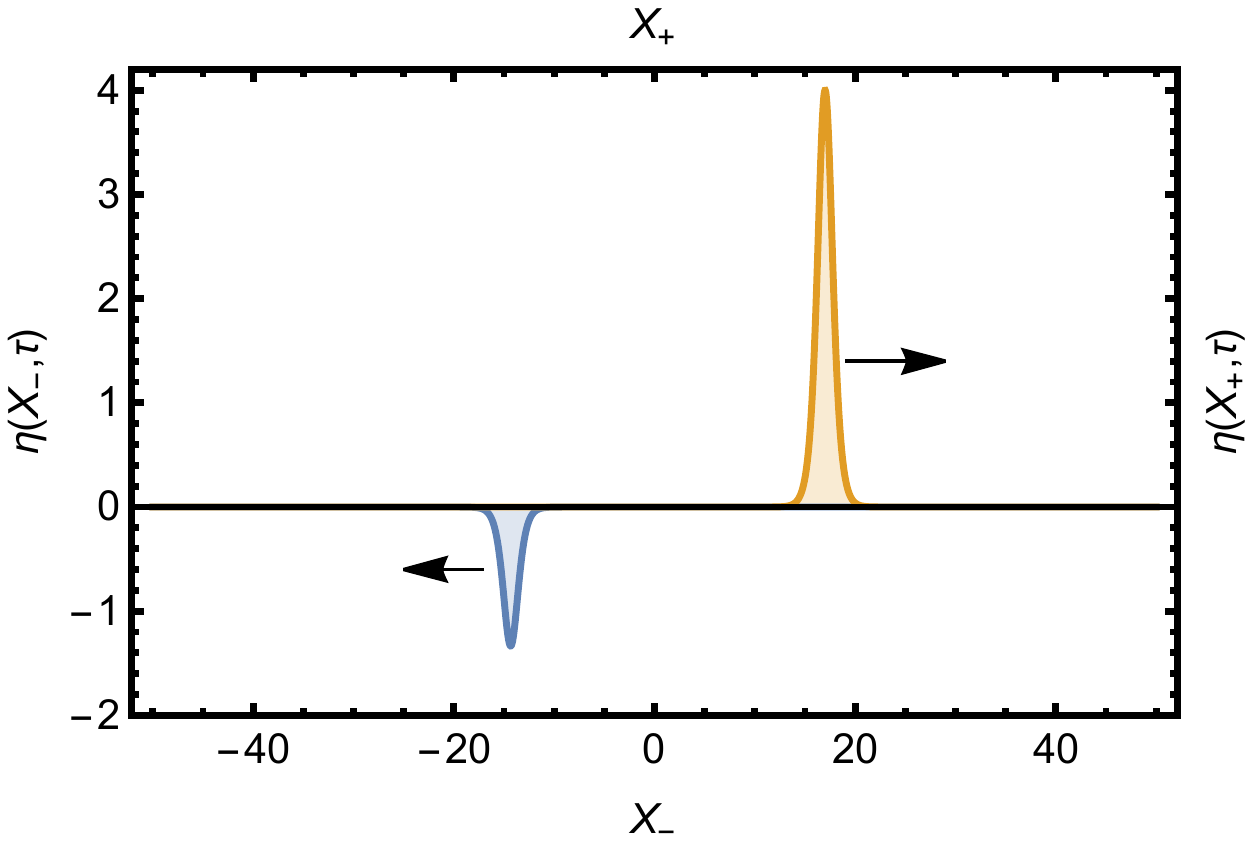}
\caption{Left (blue) and right (yellow) moving soliton in the strong parity breaking regime with parameters $|\bar \nu_o|=\frac{1}{3}$, $\bar k=1$ and $\tau=1$.}
\label{fig:sp}
\end{figure}

{\it `Strong' parity breaking regime  $|\bar\nu_o|>\tfrac{1}{6}$:} In this case, $\left(\tfrac{1}{6}\mp\bar\nu_o\right)$ have opposite signs. We call this {\it strong parity breaking regime}, because the difference between left and right moving solitons is visual, that is, one sector has positive amplitude, whereas the other corresponds to solitonic waves of depression or depletion as shown in Fig.~\ref{fig:sp}. 

\begin{figure}
\centering
\includegraphics[scale=0.7]{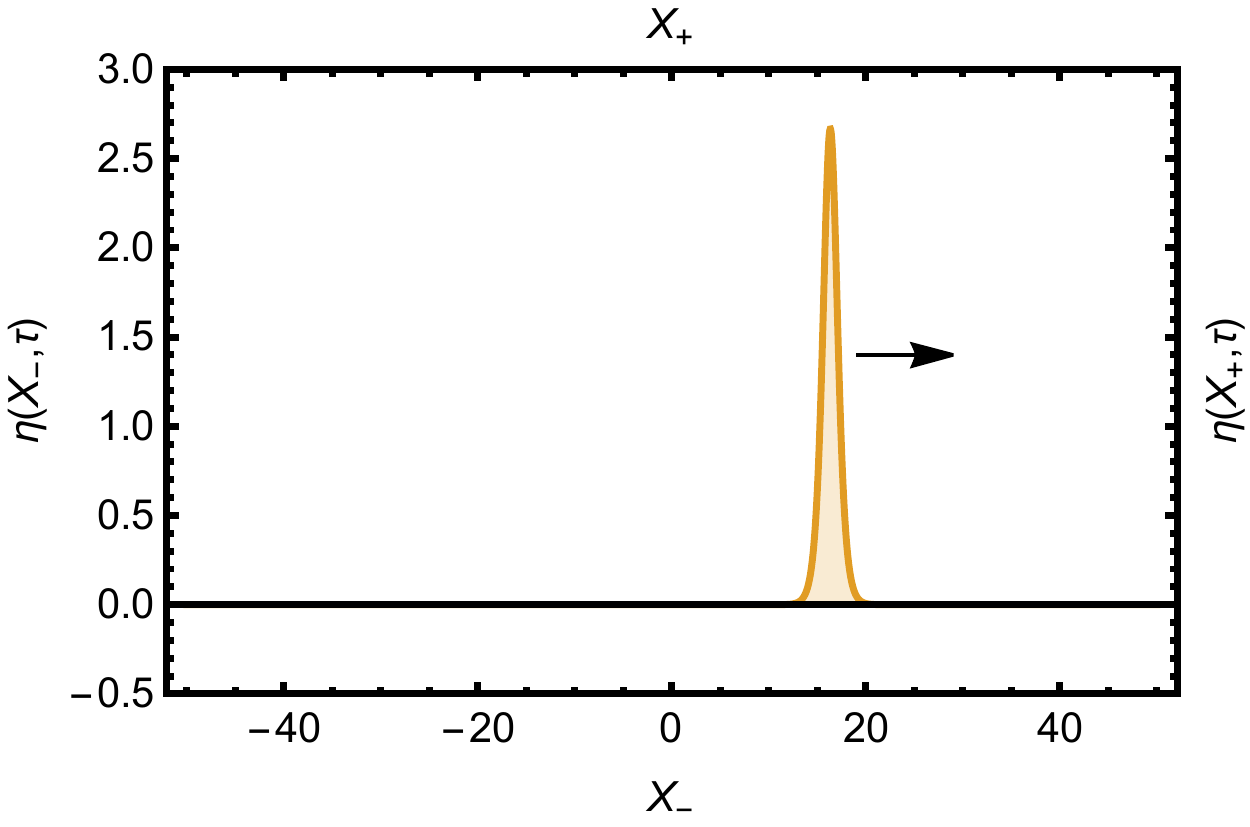}
\caption{Right (yellow) moving soliton in the critical parity breaking $|\bar \nu_o|=\tfrac{1}{6}$, $\bar k=1$, and $\tau=1$. The left mover has zero amplitude for this case.}
\label{fig:cp}
\end{figure}

{\it `Critical' dynamics $|\bar\nu_o|=\tfrac{1}{6}$:}
At this critical points, the dispersive term in one of the sectors vanish and we end up with the inviscid Burger's equation for such sector as shown in Fig.~\ref{fig:cp}. In fact, it is known that solutions of the inviscid Burger's equation are subjected to a blow up time, in which the spatial derivative of $\mathfrak y$ becomes infinite and higher order derivative terms become important.

In general, non-localized solutions for the KdV are given in terms of the Jacobi elliptic function, also known as cnoidal functions,
\begin{equation}
\mathfrak y(\tau,x_\pm)=8\bar k^2\left(\tfrac{1}{6}\mp\bar\nu_o\right)\text{cn}^2\left[\bar k x_\pm\pm4\bar k^3\tau\left(\tfrac{1}{6}\mp\bar\nu_o\right);\kappa\right].
\end{equation}
The parameter $\kappa\in [0,1]$ interpolates between the long linear waves for $\kappa\rightarrow 0$ $(\text{cn}\rightarrow \cos)$ and the single soliton solution for $\kappa\rightarrow 1$ $(\text{cn}\rightarrow \sech)$.

{\bf Discussion and Outlook:} In this letter, we derived parity broken generalization of the Korteweg de-Vries equation for shallow depth fluid with odd viscosity and gravity in the long wavelength weakly non-linear limit. The presence of odd viscosity manifests weak and strong parity breaking regimes in the two chiral sectors of the KdV dynamics. The odd viscosity term plays the role of surface tension albeit with opposite signs for the right and left movers. In future work, we aim to specialize this result to chiral active fluids, where odd viscous effects have been observed in free surface dynamics~\cite{soni2018free}. In order to make contact with experiments, we will numerically study the Cauchy initial value problem of an initial perturbation that evolves into left and right moving solitons and quantify conditions under which weak and strong parity breaking KdV dynamics emerges.   
 
\textit{\textbf{Acknowledgments.}} 
We thank Alexander Abanov and Vincenzo Vitelli for helpful discussions and suggestions about to this project. This work is supported by NSF CAREER Grant No. DMR-1944967 (SG) and partly from PSC-CUNY Award. GM was supported by 21st century foundation startup award from CCNY.

\bibliographystyle{my-refs}
\bibliography{oddviscosity-bibliography.bib}

%



\end{document}